\documentclass[structabstract]{aa}  
\usepackage{graphicx}
\usepackage{longtable}
\usepackage{txfonts}
\begin{document}
   \title{Multi-object spectroscopy of stars in the CoRoT fields II
    \subtitle{The stellar population of the CoRoT fields IRa01, LRa01, LRa02, and LRa06}
\thanks{based on observations obtained with the Anglo-Australian Telescope in program
         07B/040 and 08B/003.}}
\authorrunning{Guenther et al.}
\titlerunning{Stellar population in the CoRoT fields II}
\author{E.W. Guenther\inst{1},
             D. Gandolfi \inst{2},
             D. Sebastian\inst{1},
            M. Deleuil \inst{3},
            C. Moutou \inst{3},
            F. Cusano \inst{4}
}
   \institute{Th\"uringer Landessternwarte Tautenburg,
              Sternwarte 5, D-07778 Tautenburg, Germany\\
              \email{guenther@tls-tautenburg.de}
         \and
              Research and Scientific Support Department, 
              ESTEC/ESA, PO Box 299, 2200 AG Noordwijk, 
              The Netherlands
          \and
              Laboratoire d'Astrophysique de Marseille, 
              38 rue Fr\'ed\'eric  Joliot-Curie, 
              13388 Marseille Cedex 13, France
           \and
              INAF - Osservatorio Astronomico di Bologna, Via Ranzani 1, 
              40127 Bologna, Italy
            }
   \date{Received March 1, 2012; accepted April 30, 2012}
  \abstract 
{With now more than 20 exoplanets discovered by CoRoT, it has often
been considered strange that so many of them are orbiting F-stars, and
so few of them K or M-stars. Up to now, studies of the relation
between the frequency of extrasolar planets and the spectral types, or
masses of their host stars has been the realm of radial velocity
surveys. Although transit search programs are mostly sensitive to
short-period planets, they are ideal for verifying these
results. This is because transit search programs have different
selection biases than radial velocity surveys. To
determine the frequency of planets as a function of stellar mass, we
also have to characterize the sample of stars that was observed.}
{We study the stellar content of the CoRoT-fields IRa01,
LRa01 (=LRa06), and LRa02 by determining the spectral types of 11\,466 stars.
Nine planet-host stars have already been identified in these fields. 
Determing the
spectral types of thousands of stars of which
CoRoT obtained high-precision light-curves also 
makes possible a wide
variety of other research projects.}
{We used spectra obtained with the multi-object
 spectrograph AAOmega and derived the spectral types by using template
 spectra with well-known parameters.}
{We find that $\mathrm{34.8\pm0.7\%}$ of the stars observed by CoRoT in these
 fields are F-dwarfs, $\mathrm{15.1\pm0.5\%}$ G-dwarfs, and $5.0\pm0.3\%$
 K-dwarfs.  We conclude that the apparent lack of
 exoplanets of K- and M-stars is explained by the relatively small
 number of these stars in the observed sample.  We also show that the
 apparently large number of planets orbiting F-stars is similarly
 explained by the large number of such stars in these fields. 
 Given the number of F-stars, we would have
 expected to find even more F-stars with planets .
 Our study also shows that the difference between the sample of stars
 that CoRoT observes and a sample of randomly selected stars is
 relatively small, and that the yield of CoRoT specifically is the detection one
 hot Jupiter amongst $\mathrm{2100\pm700}$ stars. }
{We conclude that transit search programs can be used
to study the relation between the frequency of planets and the mass of
the host stars, and that the results obtained so far generally
agree with those of radial velocity programs.  }
   \maketitle
%

\section{Introduction}

Analyzing the statistical properties of extrasolar planets provides
important clues on planet-formation. Particularly important are the
relations between the properties of host stars and their planets. Up
to now, most of these studies have been carried out using radial
velocity surveys. Radial velocity surveys have the clear advantage
that they allow one to detect planets with orbital periods ranging from
less than a day to many years, but they have the disadvantage that
they are biased against rapidly rotating and active stars. This is a
problem if one aims to determine the relation between the mass of the host
stars and the frequency of planets, because most of the main-sequence
stars with masses larger than the Sun rotate fast (e.g. Reiners \&
Schmitt \cite{reiners03}).  One possible solution is to observe giant
stars.  However, giant stars do not have close-in planets and it is
therefore difficult to compare the results obtained for giant stars with
those for main-sequence stars. Nevertheless, by combining data of
main-sequence and giant stars, Johnson et al. (\cite{johnson10})
concluded that the frequency of planets increases proportionally to the
mass of the host stars (see also Johnson et al. \cite{johnson07}; Udry
\& Santos \cite{udry07}).

It would consequently be important to test these results with an
independent method.  Although transit search programs detect mostly
short-period planets, they offer this possibility.  The CoRoT-survey
(COnvection, ROtation, and planetary Transits) is particularly
suitable for this purpose. The discovery of a planet transiting an
F6V star with a $v\,\sin\,i$ of $40\pm5$ $\mathrm{km\,s^{-1}}$
(Gandolfi et al. \cite{gandolfi10}), and a planet orbiting a highly
active star (Alonso et al. \cite{alonso08}) shows that this survey is
not biased against rapidly rotating and active stars. Furthermore, the
detection of a planet with an orbital period of 95 days (Deeg et
al. \cite{deeg10}), and a planet with radius of $1.58\pm0.10\,
R_{\rm Earth}$ (L\'eger et al. \cite{leger09}; Bruntt \cite{bruntt10})
shows that CoRoT is able to detect planets in relatively long
orbit as well as planets of small radii. The statistical analysis of
the CoRoT survey thus will give us a complementary view to the results
obtained in radial velocity surveys, because it is less biased
against rapidly rotating and active stars. As an additional advantage,
CoRoT observes fields in different regions of the sky, allowing us to
find out whether the population of planets depends on the region in
the sky that is being observed.

At first glance there seems to be a huge difference between results
obtained in transit and radial velocity surveys. Out of the first 20
planet-hosting stars discovered by CoRoT, 35\% are F-stars, 55\%
G-stars, and 20\% are K-stars, with not a single one being an
M-star. If we consider the number of stars harboring planets with a
semi-major axis $\leq0.1$ AU discovered by radial velocity surveys,
7\% of them are F-stars, 39\% G-stars, 39\% K-stars, and 13\% M-stars.
It has therefore often been considered strange that CoRoT finds so
many planets orbiting F-stars, and so few orbiting K and M-stars.
However, before we can conclude that there is a difference between the
two types of surveys, we have to know how many F, G, K and M-stars the
samples contains.

To compare the results obtained in radial velocity surveys
with those obtained by CoRoT, we have to know how many F, G, and
K-type stars CoRoT has observed. This is the aim of this work.  By
spectroscopically characterizing the sample of stars that CoRoT has
observed, we answer the question whether relatively few
planets are found around K- and M-stars because of the lack of such
objects in the sample, or whether it reflects a real lack of close-in
planets around these types of stars.  In the same way, we also answer
the question why CoRoT finds so many planets of F-stars.

\section{Concept of the survey}

This work is part of a series of articles devoted to the
spectroscopic study of the stellar population in the CoRoT exoplanet
fields. Using intermediate-resolution spectroscopy acquired with the
FLAMES/GIRAFFE multi-fibre facility at ESO-VLT, Gazzano et al.
(\cite{gazzano10}) determined $v\,\sin\,i$, $V_{rad}$, $T_{\rm eff}$,
$log(g)$, [M/H], and [$\alpha$/Fe] for 1227 stars in the CoRoT-fields
LRa01, LRc01, and SRc01. While 1227 stars sounds like a high number, 
it means that only $\sim4$\% of the stars that CoRoT observed in these fields
were analyzed. Additionally, the survey contained stars in fields
located in two opposite directions in the sky (the so-called galactic
"center" and "anti-center"-eyes of CoRoT), which makes the sample of
stars for each "eye" relatively small.  There is consequently a need for a
more comprehensive study in which a larger sample of stars is
analyzed.

There are more potential target stars in the CoRoT exoplanet fields
than there are stars observed by the satellite.
Based on a massive $UBVr'i'$ photometric survey performed during the
mission preparatory phase (Deleuil et al. \cite{deleuil09}), the
CoRoT-targets were selected by giving higher priority to
photometrically identified dwarf stars. It is thus also important to
find out how much the CoRoT-sample differs from a sample of randomly
chosen stars. To quantify the effect of the selection
process, we analyzed 7131 stars observed by CoRoT and 4335 stars that
were not observed by CoRoT but have the same brightness and are in the
same regions as the CoRoT target stars.

The results of our spectroscopic survey are published in two
articles.  In paper I (Sebastian et al. \cite{sebastian12}), we gave a
list of stars with spectral types O, B, and A in IRa01, LRa01, and
LRa02.  Here we analyze stars with spectral types F, G, K, and M.
The LRa01-field was recently re-observed by 
CoRoT, the field is called LRa06 in the new observations.

\section{Spectroscopic observations and data reduction}

Observations, data-reduction, and spectral type determination are
described in detail in paper I. For convenience, a short
summary is provided below.

We used the AAOmega multi-object spectrograph mounted on the
Anglo-Australian telescope during two observing campaigns, i.e., from
13 to 20 January 2008 and from 28 December 2008 to 4 January
2009. AAOmega is ideal for our purpose, because it has a field of view
of about two degrees, which matches the CoRoT-fields IRa01,
LRa01, and LRa02 well, which have a size of $1.8\degr \times 3.6\degr$ . 
In the blue arm we employed the \emph{580V}-grating, which covers the
wavelength range 3740-5810~\AA , and in the red arm the
\emph{385R}-grating, which covers the 5650-8770~\AA \, range. The
spectral resolution of both gratings is $\lambda/\Delta\lambda \sim
1300$. Using the AAOmega data-reduction pipeline \emph{2dfdr}
(Saunders et al. \cite{saunders04}; Smith et al. \cite{smith04}), we
subtracted the bias from the frames, flat-fielded them, and extracted
and wavelength-calibrated the spectra.  The sky-background was then
subtracted from the spectra using the data obtained with fibres that
were placed on sky-background. In the last step we flux-calibrated the
spectra. In total we obtained spectra for 11\,466 stars.

\begin{figure}
  \includegraphics[height=.30\textheight]{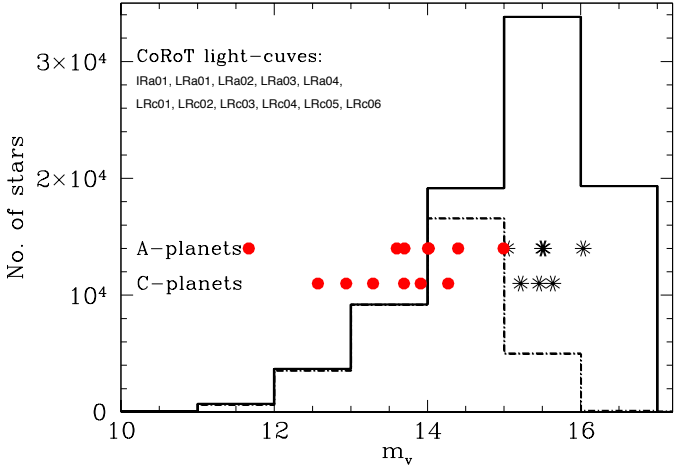}
\caption{Histogram of the visual magnitudes of the stars observed by
         CoRoT in eleven exoplanet fields.  The full line represents
         all stars, the dashed line only those that were observed in
         the CHR-mode. The thick (red) points indicate the brightness
         of the planet-host stars found in the "galactic center"
         (C-planets) and "anti-center eyes" (A-planets) of CoRoT.  The
         asterisks shows the brightness of the planet-hosting stars
         observed in the MONO-mode.}
  \label{histLC}
\end{figure}

Fig.\,\ref{histLC} shows the histogram of the visual magnitudes of the
stars observed by CoRoT in eleven exoplanet fields. CoRoT is equipped
with a bi-prism that disperses the light into a short spectrum. For
relatively bright stars, the flux is measured in three colors
(CHR-mode), for faint stars only the total flux is determined
(MONO-mode). The dashed line in Fig.\,\ref{histLC} gives the number of
stars per magnitude interval monitored in the CHR-mode, and the full
line the number of stars observed with the CHR and the MONO-mode
together.  Down to $V\sim 15$~mag essentially all stars are observed
in the CHR-mode. To focus on stars that were observed in
this mode, we limited our survey to $V\le15$ mag. We focused on
stars observed in the CHR-mode, because the quality of this data-set
is higher.  For example, from the first 20 planet host stars
discovered by CoRoT, seven were discovered with the MONO-mode, and 13 with
the CHR-mode.  The points in Fig.\,\ref{histLC} represent the
brightness of the planet host stars observed in the CHR-mode, and the
asterisks the planet host stars observed in the MONO-mode.

In total, CoRoT observed 3097 stars in IRa01, 7470 stars in LRa01
(LRa06), and 4125 stars in LRa02 with the CHR-mode.  This means
that we determined the spectral types of more than 50\% of the stars
observed by CoRoT in the CHR-mode in these fields. We also
  obtained sufficient spectra of stars that were not observed by CoRoT,
  which allows us to find out by how much the stellar population of
  the CoRoT-targets differs from the general population of stars in
  these fields.  The spectral types were derived by iteratively
fitting the observed spectra with templates taken from a library of
spectra.

\section{The accuracy of the spectral type determination}

\begin{figure}
  \includegraphics[height=.25\textheight]{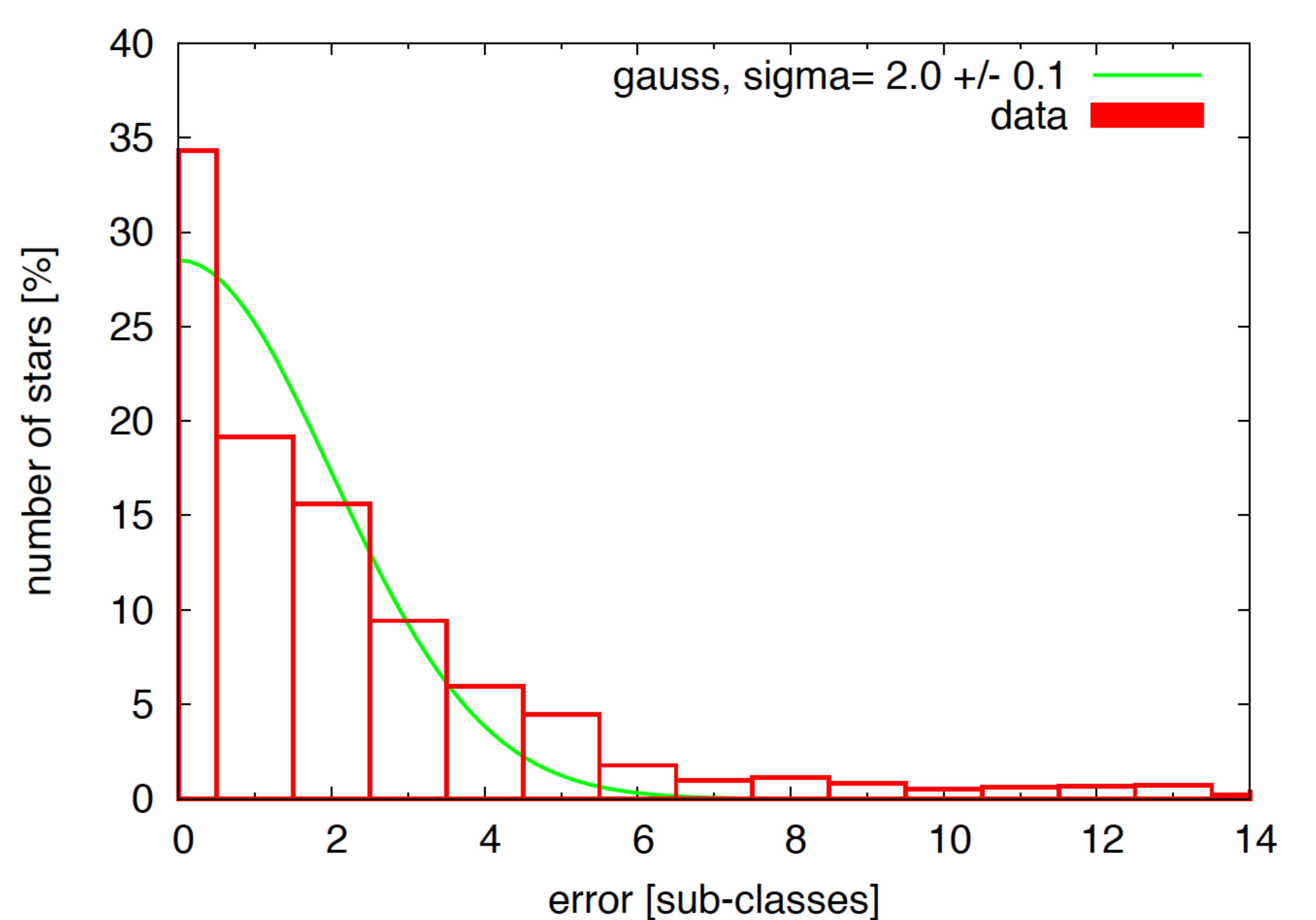}
  \caption{Histogram of the spectral type differences. The error 
             on the spectral class determination is
             on average $2.0\pm0.1$ sub-classes.}
  \label{errorSpec}
\end{figure}

For 33\% of the stars observed with AAOmega, we obtained more than one
spectrum. This strategy not only increases the accuracy of the
spectral type determination but it also allowed us to assess the
internal precision of our method by comparing the spectral types
derived for the same star using different
spectra. Fig.\,\ref{errorSpec} shows the histogram of the difference
of the spectral classifications obtained for two or more spectra of
the same star.  The error is on average $2.0\pm0.1$ sub-classes. If we
perform the same analysis for the different spectral-classes, we find an
error of $1.90\pm0.08$ sub-classes for B-stars, $1.28\pm0.07$
sub-classes for A-stars, $2.2\pm0.1$ sub-classes for F-stars,
$2.9\pm0.2$ sub-classes for G-stars, and $1.7\pm0.1$ sub-classes for K-
and M-stars.  As shown in Fig.\,\ref{errorSpec}, there are a few stars
for which the difference is larger than five sub-classes. These
outliers are caused by intrinsic variability of the stars, binaries,
or by instrumental effects, like spectra with very low signal-to-noise
ratio. Although we include spectra of very low signal-to-noise ratio
in Fig.\,\ref{errorSpec}, we did not use such spectra for the
classification.  The true error therefore is certainly smaller than two
sub-classes.
 
However, because in our method we essentially compare the observed
spectra with templates, the quality of the templates and the accuracy
with which their spectral types were determined is important. When
selecting the library of spectra, there are three aspects to be taken
into account: (1) The library of spectra should cover all spectral
types and luminosity classes that are of interest in this work. (2)
The spectral types given for the templates should be accurate. (3) The
library must contain templates with a resolution that is at least as
high as that of AAOmega. The three most promising libraries are
Valdes et al. (\cite{valdes04})(Val), Jacoby, Hunter \& Christian
(\cite{jacoby84})(JHC), and Le Borgne (\cite{borgne03})(STELIB).

Our method does not derive the spectral types of the stars
directly but identifies which spectrum from a library matches the
observed spectrum best.  The spectral type of the CoRoT-target is then
given by the spectral type of that template. If we take for example
CoRoT102899501 (=IRa01\_E2\_649), we find that HD 22468 (Val), HD
29050 (JHC), and HD 132141 (STELIB) match the observed spectrum best.
The spectral types derived are thus G9V (Val), G9V (JHC), and K1V
(STELIB), respectively.  However, instead of using the spectral types
of these stars given in these three articles, we can also look up
their spectral types in Skiff (\cite{skiff10}). Skiff (\cite{skiff10})
compiled a long list of spectral-types of many stars from the
literature. For HD 22468 Skiff (\cite{skiff10}) lists three spectral
types: K2Ve, G8Ve and K2:Vn\,k, and for HD 132141 K0V and K1V. HD
29050 is not listed in in this work but SIMBAD gives K1V. The values
published in the literature for these stars generally agree
with the values given in the three articles.

The best test of our method and the quality of the templates would be
if we were to know the exact spectral type for some stars that we
observed with AAOmega. Based on high-resolution spectra, 
Gondoin  et al. (\cite{gondoin12}) determined for
CoRoT102899501 $T_{\rm  eff}=5180\pm80$ 
and $4.35\pm0.10$. Using the the conversion from $T_{\rm eff}$ and 
$\log\,g$ into spectral type given in Gray  (\cite{gray09}) this corresponds 
to a K1V star. The values derived from the AAOmega-spectra thus differ by $\leq $ 
2 sub-classes from the true value. This is consistent with the internal 
error shown in Fig.\,\ref{errorSpec}.

We can carry out similar tests using the planet-host stars, because
their stellar parameters have been determined with very high accuracy
using high signal-to-noise spectra obtained with UVES and HARPS.  Our
AAOmega-survey contains the planet-host stars \object{CoRoT-1} (Barge
et al. \cite{barge08}), \object{CoRoT-5} (Rauer et
al. \cite{rauer09}), and \object{CoRoT-7} (L\'eger et
al. \cite{leger09}; Bruntt et al. \cite{bruntt10}).  Using the three
libraries we find for G0V-star \object{CoRoT-1} the spectral types F6V
(Val), F4III (JHC), and G0V (STELIB), respectively. For the F9V-star
\object{CoRoT-5} the three different templates give F9IV (Val), F7V
(JHC), and F7V (STELIB), and for the G9V-star \object{CoRoT-7} G8V
(Val), G9V (JHC), K1V (STELIB).

If we consider that the resolution of the AAOmega-spectra is
too low to distinguish luminosity-class IV from V stars, we obtain the
correct luminosity class in all cases, except for the library
published by Jacoby, Hunter \& Christian (\cite{jacoby84}) in the case
of \object{CoRoT-1} . The differences between the true spectral types
and the spectral types derived from the AAOmega spectra are consistent
with the internal error.  Since the library of spectra published by
Valdes et al. (\cite{valdes04}) covers all spectral types and
luminosity classes, and since the stars in this library have recently
been precisely re-calibrated by Wu et al. (\cite{wu11}), we used these
templates.

\begin{table*}
\caption{Frequency of stars in IRa01, LRa01, LRa02}
\begin{tabular}{l c r r r r r r r }
\hline
status & luminosity & O & B & A & F  & G  & K  & M \\
         & class        & [\%] & [\%] & [\%] & [\%] & [\%] & [\%] & [\%] \\ 
\hline
stars observed   & IV+V & $<0.01$ & $3.9\pm0.2$ & $15.9\pm0.5$ & $34.8\pm0.7$ & $15.1\pm0.5$ & $5.0\pm0.3$ & $<0.01$ \\ 
with CoRoT       & I,II,III & $<0.01$ & $1.5\pm0.1$ & $3.0\pm0.2$ & $4.8\pm0.3$ & $9.6\pm0.4$ & $6.0\pm0.3$ & $0.3\pm0.06$ \\ 
\hline
stars not observed & IV+V & $0.02\pm0.02$ & $4.7\pm0.3$ & $18.9\pm0.7$ & $27.7\pm0.8$ & $13.9\pm0.6$ & $4.6\pm0.3$ & $<0.02$ \\ 
with CoRoT            & I,II,III & $0.02\pm0.02$ & $1.5\pm0.2$ & $2.8\pm0.3$ & $4.8\pm0.3$ & $11.1\pm0.5$ & $9.1\pm0.5$ & $0.8\pm0.1$ \\ 
\hline
all                      & IV+V & $0.01\pm0.01$ & $4.2\pm0.2$ & $17.0\pm0.4$ & $32.1\pm0.5$ & $14.6\pm0.4$ & $4.9\pm0.2$ & $<0.01$ \\ 
stars                  & I,II,III & $0.01\pm0.01$ & $1.5\pm0.1$ &   $3.0\pm0.2$ &  $4.7\pm0.2$  & $10.2\pm0.3$ & $7.2\pm0.3$ & $0.6\pm0.1$  \\
\hline
\end{tabular}
\label{tab01}
\end{table*}

\section{Results}

The primary result of the our survey is the determination of the
spectral types of 11\,466 stars in CoRoT-fields IRa01, LRa01 (LRa06),
and LRa02.  In Table 3 we give the spectral types of F, G, K, and
M-stars in these fields, and in paper\,I (Sebastian et
al. \cite{sebastian12}) the spectral types of all O, B, and
A-stars. For completeness we list in column 7 and 8 in Table 3 the
$T_{\rm eff}$ and $\log\,g$-values for the stars that were also
observed by Gazzano et al.  (\cite{gazzano10}). Knowing the
spectral types of thousands of stars for which CoRoT obtained
high-precision light-curves in three colours opens up 
possibilities for a wide
variety of research projects. For example, we now determined the
spectral types of many stars discussed in Kabath et
al. (\cite{kabath07}) Kabath et al. (\cite{kabath08}), and Kabath et
al. (\cite{kabath09}).

The first application of our results was to calculate the planet-yield
of CoRoT. In total 21\,280 light curves of IRa01, and LRa01 have been
analyzed (Carpano et al. (\cite{carpano09}) and Carone et
al. (\cite{carone11}), and 91 candidates were found, of which at least
six are hosting planets. Five of these are hosting hot Jupiters
($0.45\,M_{Jup}$ $\le$ $M_p$ $\le$ $2.5\,M_{Jup}$; $P<10\,d$).  Since
all of the known hot Jupiters in these fields orbit F- and G-dwarfs,
and since $49.9\pm0.8\%$ of the stars in the sample are stars of this
type, the yield of CoRoT is one hot Jupiter amongst $2100\pm700$
stars. If we take the geometric probability for a transit into
account, the true frequency of hot Jupiters is $0.4\pm0.2\%$. This
result agrees well with the results obtained in radial velocity
surveys. For example, Cumming et al. (\cite{cumming08}) and Naef et
al. (\cite{naef05}) reported frequencies of $0.4\pm0.3\,\%$
($P<11.5\,\mathrm{d}$), and $0.7\pm0.5\,\%$ ($P<5\,\mathrm{d}$),
respectively.  The fact that the number of hot Jupiters detected with
CoRoT agrees with the number of hot Jupiters discovered in
radial velocity surveys implies that CoRoT detects all transiting
planets of this type in these fields.

As mentioned above, when selecting the CoRoT targets, preference was
given to stars that were more likely to be late-type dwarfs.  With the
spectral types derived, we can now asses the effect of the selection
process.

\begin{figure}
  \includegraphics[height=.28\textheight]{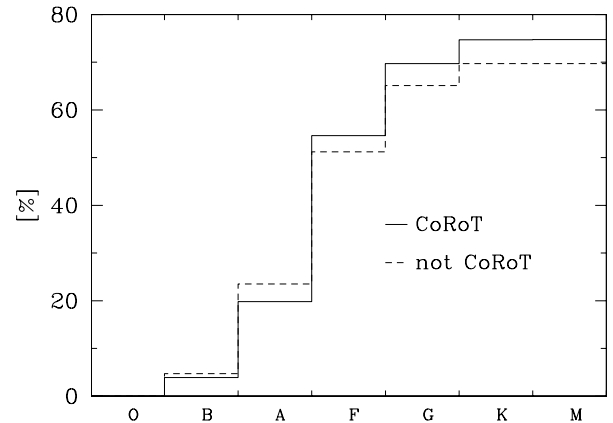}
  \caption{Cumulative distribution function of dwarf stars that were
                observed and not observed with CoRoT.}
  \label{cumulativeAll}
\end{figure}

Fig.\,\ref{cumulativeAll} shows the cumulative distribution function
for dwarf stars observed, and not observed by CoRoT\footnote{The
  cumulative number is smaller than 100\%, because not all stars are
  dwarfs}. A Kolmogorov-Smirnov test shows that the samples are indeed
different.  The main difference is that the CoRoT-sample contains
more F- and G-dwarfs than the sample of stars that was not observed by
CoRoT.  This is obviously the effect of the selection process of the
targets. The difference of the stellar populations of the two samples
is small however. This can also be seen in Table \ref{tab01}, were
we list the frequency of stars for each spectral type for the two
samples.  For example, the fraction of dwarfs (sub-giants and dwarfs)
is $75.3\pm1.0\%$ for the CoRoT sample, compared to $69.9\pm 1.3\%$
for the sample of stars that was not observed by CoRoT. The fraction
of O and B stars in the two samples is $5.4\pm0.3\%$, and
$6.2\pm0.5\%$, respectively.  The stellar population of the
CoRoT-sample thus is not that different from the general population of
stars in these fields.

The distribution of spectral types in a sample of stars depends on the
region in the sky that is being observed and on the limiting magnitude
of the survey. If we were, for example, to carry out a deep survey at a
high galactic latitude, we would find only very few intrinsically
bright stars, because such stars would have to be at a large distance
to be that faint but the number of stars decreases with the
distances from the plane of the milky way (see for example Phleps et
al. \cite{phleps05}). If we observe at low galactic latitude,
extinction complicates somewhat more the picture but the
fraction of intrinsically luminous stars will still decrease for
fainter stars.  For example, Robin et al. (\cite{robin03}) showed that
giants dominate for $V<11$, and main-sequence stars for
$V>14$. Because the sample in this study contains stars from V=10 to
V=15, we expect that it will contain a relatively high percentage
of late-type main sequence stars. A high percentage of late-type 
main-sequence stars is an advantage for a planet search program.
  
Prior to the CoRoT-observation, the fields IRA01, LRa01, and LRa02
were observed photometrically in the $U,B,V,r',i'$ bands.  After
combining these data with 2MASS-photometry, the photometric spectral
types of the stars were determined. When selecting the CoRoT-targets,
preference was given to stars that had a higher probability of being
late-type dwarfs, and to stars with a low content of contaminating
stars. All information was made public in the form of the
EXODAT-catalog (Deleuil et al. \cite{deleuil09}). The CoRoT-sample
therefore is not a random sample of stars but a sample that was optimized
for planet search.  Using our results, we can now quantify what
fraction of stars in these fields are suitable stars for a planet search
(e.g. A- to M-dwarfs).

\begin{figure}
  \includegraphics[height=.27\textheight]{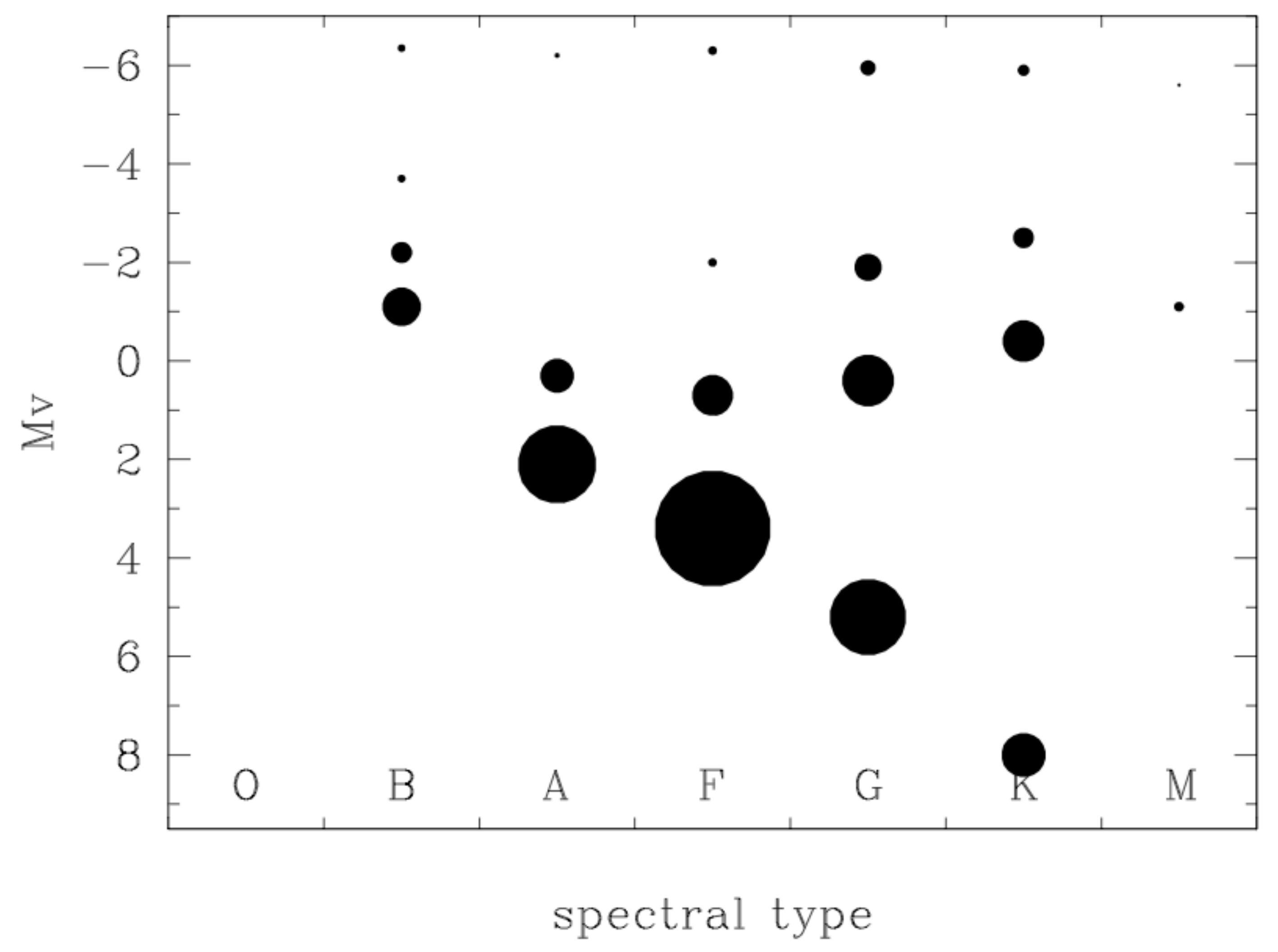}
  \caption{HRD of the stars observed by CoRoT. The size of the circles
               is proportional to the number of stars of that
               category.}
  \label{HRDCoRoT}
\end{figure}

\begin{figure}
  \includegraphics[height=.25\textheight]{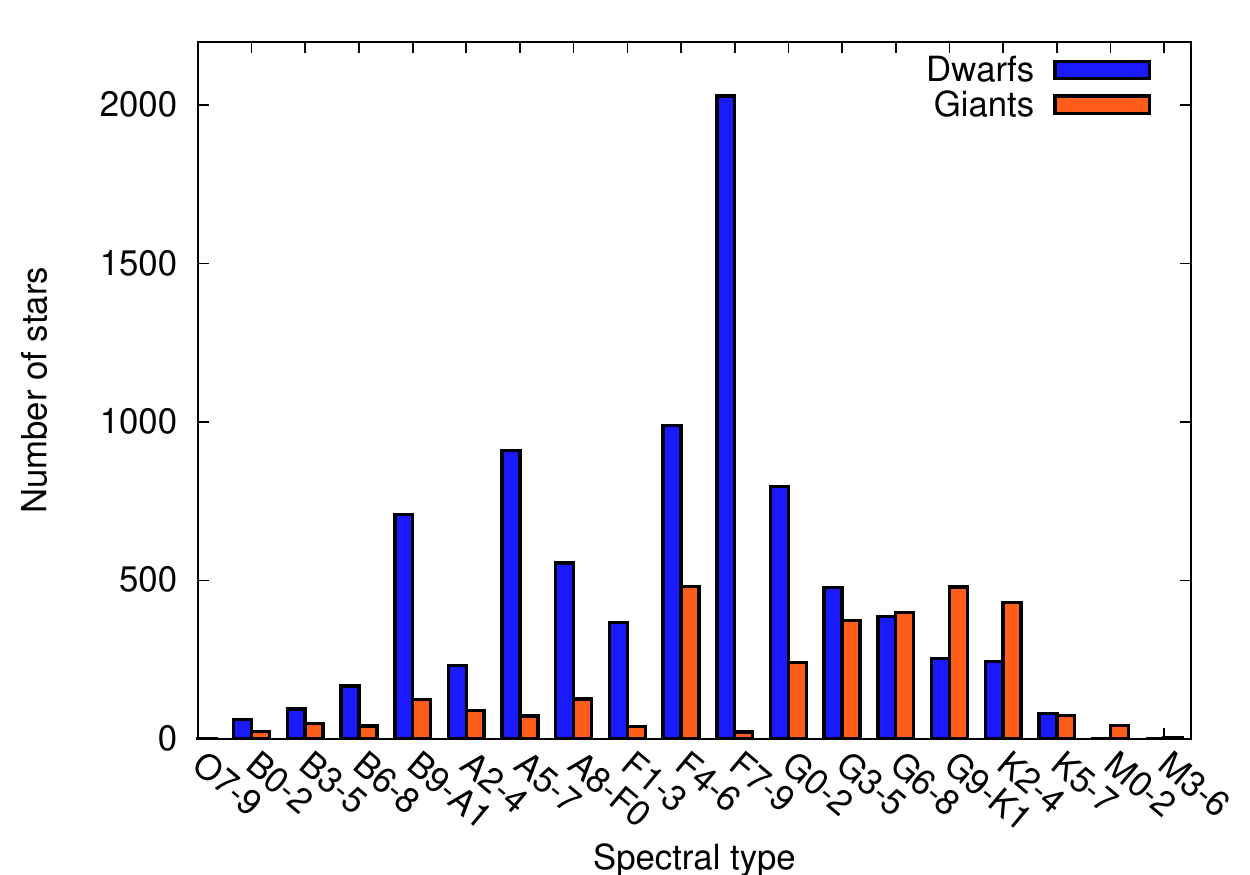}
\caption{Distribution of spectral types in our sample. The blue column
   is the sum of sub-giants and dwarfs (designated as dwarfs), and the
   red column the sum of super-giants, bright giants and giants
   (designated as giants).  The most common species of stars are late
   F-dwarfs.}
  \label{histogramSpec}
\end{figure}

Fig.\,\ref{HRDCoRoT} shows the distribution of CoRoT-targets in the
Hertzsprung-Russell diagram (HRD). The area of each of the circles is
proportional to the number of stars of that
class. Fig.\,\ref{HRDCoRoT} clearly shows that the most abundant type
of stars in the CoRoT-sample are F-dwarfs. Fig.\,\ref{histogramSpec}
shows the distribution of the spectral types in more detail.  For this
figure we added up stars of luminosity class IV and V, which we called
``dwarfs'', and all stars with luminosity class I, II, and III, which
we called ``giants''.  It turns out that the most abundant type of stars
are late F-dwarfs.  M-stars are rare amongst the CoRoT-targets, and
most of them are giants, not dwarfs.

\begin{figure}
  \includegraphics[height=.27\textheight]{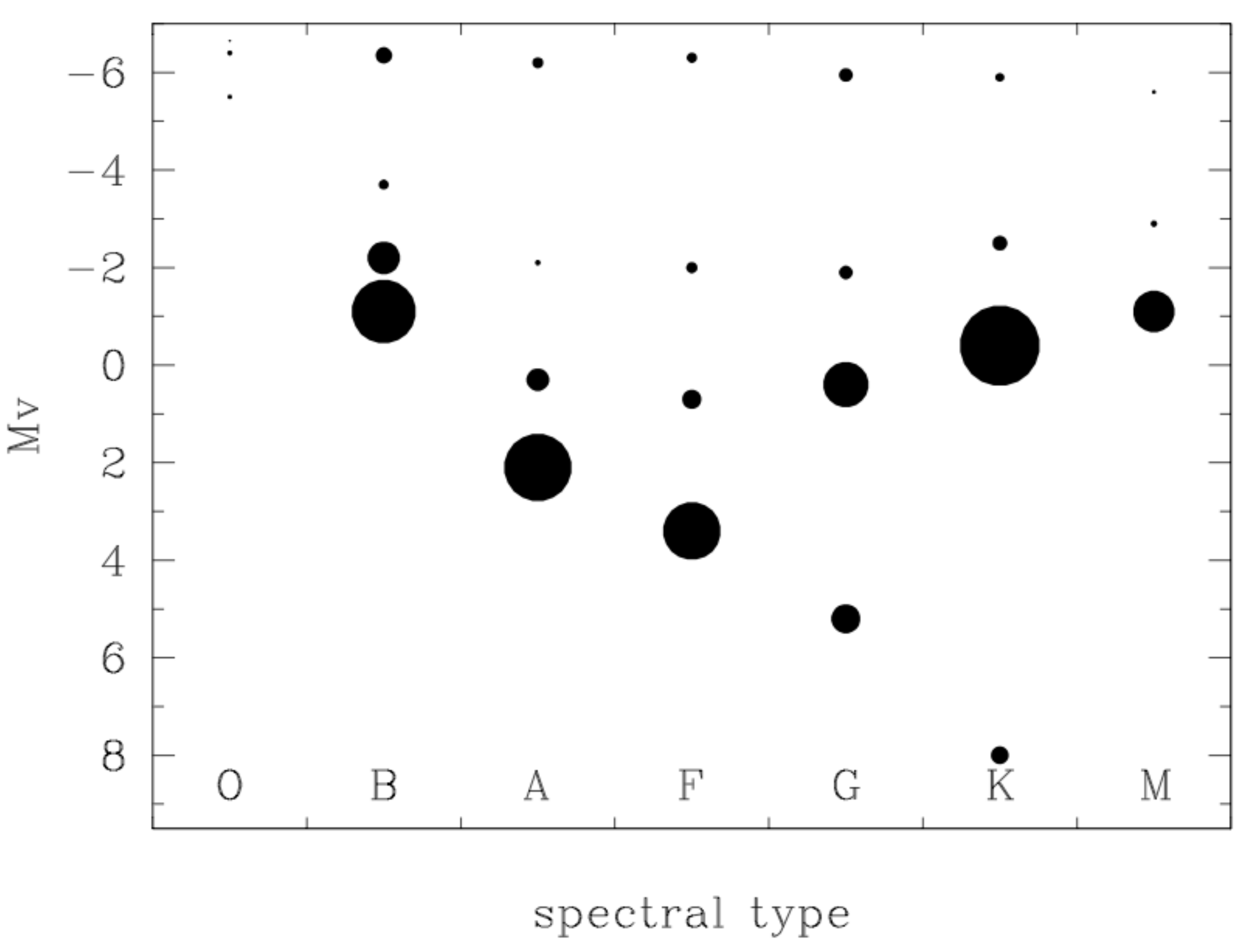}
  \caption{Same as Fig.\,\ref{HRDCoRoT} but for stars brighter
                than 6.5 mag.}
  \label{HRDbright}
\end{figure}

Fig.\,\ref{HRDbright} shows the same kind of diagram for bright stars
for comparison.  We selected for Fig.\,\ref{HRDbright} stars brighter
than V=6.5~mag so that the number of stars in Fig.\,\ref{HRDCoRoT} and
Fig.\,\ref{HRDbright} is about the same. The CoRoT-sample contains
more A, F, and G-dwarfs than the bright-star sample. We find that the
fraction of suitable targets for a planet-search program is
$73.6\pm1.0\%$ for CoRoT, compared to $28.6\pm0.6\%$ for a bright-star
sample. This shows to which degree the CoRoT-sample is optimized for
planet-detection.

\section{Discussion and conclusions}

Of these nine planet-host stars discovered in the three CoRoT-fields
studied in this work, four are F-type stars (\object{CoRoT-4}, Moutou
et al. \cite{moutou08}; \object{CoRoT-5}, Rauer et al. \cite{rauer09};
\object{CoRoT-14}, Tingley et al. \cite{tingley11}; \object{CoRoT-21}
P\"atzold et al. \cite{paetzold11}), four are G-type stars
(\object{CoRoT-1}, Barge et al. \cite{barge08}; \object{CoRoT-7},
L\'eger et al. \cite{leger09}; \object{CoRoT-12}, Gillon et
al. \cite{gillon10}; \object{CoRoT-13}, Cabrera et
al. \cite{cabrera08}), and one is a K-star (\object{CoRoT-24}; Alonso
et al. \cite{alonso12}). With only nine planet-host stars, the sample
is too small to answer the question whether or not
the number of planets increases with the mass of the host star.
 
However, we can still discuss whether the high percentage
of planets orbiting F-stars and the small fraction orbiting K-stars 
can be explained with the observed sample of stars, 
or whether it can already be taken as evidence for
an increase of the planet frequency with the mass of the host
star.

Given that only $5.0\pm 0.3\%$ of the stars in the sample are K- and
M-dwarfs, the small number of planets found in K- and M-stars by CoRoT
is due to the lack of K- and M-stars in the sample.  The situation is
more complicated for the F-stars. Since four of the planet-hosting
stars are G-stars, and there are 2.3 times as many F- than G-stars,
there should have been 9 to 14 planet-hosting F-stars, depending on
whether we assume that the frequency of planets increases with the
mass of the star, or not. In contrast to this, only four
planet-hosting stars of this type were found. 
Of course, this is still low-number statistics, but we can conclude 
that there is no excess of close-in F-star planets.  
Quite contrary to this, we would have expected to find even
more planets of F-stars.

If we add up the results obtained for all galactic anti-center fields,
CoRoT has found four F-stars and eight G-stars that are hosting
planets. If there were also twice as many F- as G-stars in these
fields, we would have expected to find four times as many F-stars hosting
planets than we have found.  However, we have to carry out a similar
study of all the other fields before we can draw any firm
conclusion. In any case, our study shows that the statistical analysis
of transit search programs can lead to very interesting results. Once
the CoRoT survey is completed, it will be possible to answer the
question whether or not there is a lack of close-in planets of F-stars.

Up to now, we have only analyzed stars located in the so-called
"galactic anti-center region" of CoRoT. The next step would be to
carry out a similar survey for the fields in the so called "galactic
center eye" of CoRoT. Apart from improving the statistics, there is
another good reason: Because of the metallicity gradient in our
galaxy and the correlation between planet frequency and metallicity,
it is expected that stars orbiting closer to the galactic center
should have a higher frequency of planets than stars orbiting at
larger distances.  If we take the gradient of
$-0.06\pm0.01\,\mathrm{kpc^{-1}}$ from Friel et al. (\cite{friel02})
and the relation between metallicity and planet frequency from Santos
et al. (\cite{santos04}), we expect a difference of a factor of two
for the planet frequency of stars with galactocentric distances that
differ by 2 kpc. Using a more sophisticated model, Reid
(\cite{reid08}) estimated that the difference of the planet-frequency
is a factor of 1.8 for stars with galactocentric distances between 7
and 9 kpc. Since CoRoT observes stars in this range CoRoT-data might
show this effect, if all fields are analyzed. Interestingly, the
galactocentric distance from 7 to 9 kpc corresponds also to the
galactic habitable zone (Lineweaver \cite{lineweaver04}).  By
extending the survey to the "galactic center" fields we will thus not
only improve the statics but we might even be able to find out whether
the properties of the planets depend on the galactocentric distance.

\begin{acknowledgements}

We are grateful to the user support group of AAT for all their help
and assistance for preparing and carrying out the observations.  We
would like to particularly thank Rob Sharp, Fred Watson and Quentin
Parker. This research has made use of the SIMBAD database, operated at
CDS, Strasbourg, France.
  
\end{acknowledgements}

\onecolumn
\begin{center}
\begin{longtable}{l l r l c c c}
\caption{Late type stars in IRa01, LRa01, and LRa02} \label{tab03} \\
\hline
CoRoT-id & RA        & DEC       & V        & Spec         & $T_{\rm eff}$ & $\log\,g$ \\
               &(2000.0) & (2000.0) &[ mag] & AAOmega & [K]       & [$cm\,s^{-2}$]\\
\hline
\endfirsthead

\multicolumn{4}{c}%
{{ \tablename\ \thetable{} -- continued}} \\
\hline 
CoRoT-id & RA        & DEC       & V        & Spec         & $T_{\rm eff}$ & $\log\,g$ \\
               &(2000.0) & (2000.0) &[ mag] & AAOmega & [K]       & [$cm\,s^{-2}$]\\
\hline
\endhead

\endfoot

\hline \hline
\endlastfoot
102656576 & 100.685898   & -1.72535   & 11.736   & F/Mcomp & & \\
102741522 & 101.146553 & 0.28083 & 10.468 & F0Ib & & \\
102968255 & 102.47985 & -3.01884 & 13.946 & F0II & & \\
102567769 & 100.130638 & 0.60634 & 11.203 & F0II & & \\
102809733 & 101.568489 & -2.60183 & 13.753 & F0III & & \\
110750692 & 102.81175 & -5.41552 & 14.002 & F0III & & \\
110839159 & 102.71896 & -3.96915 & 14.198 & F0IV & & \\
102726045 & 101.059143 & -3.21953 & 13.104 & F0IV & & \\
\end{longtable}
\end{center}
\twocolumn

\end{document}